\documentclass{elsarticle} 
\usepackage{amsmath,mathtools,amssymb}
\allowdisplaybreaks%允许公式分页
\usepackage{graphicx}  
\usepackage{subfigure}
\usepackage{color}
\usepackage{verbatim} 
\usepackage{booktabs}
\usepackage{bm}  
\usepackage{dutchcal}
\usepackage[left=3cm,right=3cm,top=2.5cm,bottom=2.5cm]{geometry} 
\usepackage{lineno,hyperref}
\usepackage{tablists} 
\usepackage{makecell} 
\usepackage{tikz}
\usetikzlibrary{shapes.geometric, arrows.meta, positioning, fit, backgrounds}
\usepackage{amsmath, amssymb}
\usepackage{hyperref}
\usepackage{pstricks}
\usepackage[titletoc]{appendix}
\usepackage{pstricks-add}
\usepackage{float}
\usepackage{mathrsfs} 
\usepackage{xcolor}
\usepackage{url}
\biboptions{numbers,sort&compress}
\numberwithin{equation}{section}

\journal{Elsevier} %指出投稿的期刊

\begin{document}
    \begin{frontmatter} %开始组织Front Matter

        \title{Higher Wess–Zumino–Witten term from semistrict higher Chern–Simons theory} %论文题目
        
        \author{Mengyao Wu\corref{cor1}}
        \ead{mengyao_wu_w@163.com}
        \address{School of Mathematical Sciences, Capital Normal University, Beijing, China}
        
        \date{}
     \begin{abstract}
         We show that in any $2n+2$ dimensions, the higher Chern–Simons action built from a semistrict Lie 2‑algebra gives a non‑trivial higher Wess–Zumino–Witten (WZW) term under a higher gauge transformation. The key point is that the non‑zero Jacobiator directly generates the higher WZW term, whereas it vanishes when reduced to the strict case.
        
     \end{abstract}

    \end{frontmatter}

    %%\pacs[JEL Classification]{D8, H51}
    
    %%\pacs[MSC Classification]{35A01, 65L10, 65L12, 65L20, 65L70}
    \section{Introduction}
    
  The Wess--Zumino--Witten (WZW) term first arose from the study of anomalous Ward identities in the early 1970s, and was later shown to obey a topological quantization condition while unifying perturbative and non-perturbative anomalies in current algebra~\cite{WessZumino1971, witten1983global}. 
  A deeper connection emerged from ordinary three dimensional Chern--Simons theory on a manifold with boundary. Under a gauge transformation, the CS action changes by the WZW action as a boundary term.~\cite{ordwzw}. 
  This link shows that the WZW term not only captures topological features in two dimensions but also governs quantum anomalies and winding number in three dimensional gauge theory, with its coupling constant quantised to ensure path integral invariance.
  
    In the context of describing gauge interactions of extended objects (strings, branes, etc.), higher gauge theory based on categorified algebraic structures such as Lie 2-algebras and Lie 2-groups provides a natural higher-dimensional framework \cite{baze2004higherv,baze2004highervi,lada1993sh,lada1995strongly}. Within this framework, higher Chern–Simons theories \cite{danhua2023,danhua2024higher,zucchini4-cs,zucchiniopei,zucchiniopeii,zucchiniaksz} generalize ordinary Chern-Simons theory and play important roles in topological field theories and the description of self-dual higher gauge fields.
    Extending the role of the ordinary WZW term to this higher dimensional framework, the key question is whether a higher WZW term appears naturally from the variation of the higher Chern–Simons action itself.
    
  Two results have been observed:
   \begin{itemize}
        \item Ref.~\cite{zucchiniwilson, schenkel5d} considered \textit{strict} higher gauge theory based on a crossed module in four dimensions and found that the gauge variation contains no higher WZW term. Ref.~\cite{danhuawzw} extended this result to arbitrary dimension $2n+2$, proving that under a higher gauge transformation, the higher Chern--Simons action does \textit{not} generate a higher WZW term for any $n\ge 1$. The strictness forces the failure of the Maurer-Cartan equation to be trivial in a cohomological sense, and the variation remains a pure boundary term without an extra WZW piece.
       
       \item In contrast, Ref.~\cite{zucchini4-cs,soncini2014} worked in four dimensions with a \textit{semistrict} Lie 2-algebra. They found that the gauge variation does contain a higher WZW term, and the non-trivial Jacobiator directly gives rise to that term.
   \end{itemize}
   
  Thus, the four dimensional semistrict case does give rise to a higher WZW term, while the arbitrary dimensional strict case does not.
   What happens in arbitrary dimension $2n+2$ with a semistrict Lie 2-algebra?  
   This natural gap has remained open.
  
In this short note, we show that for any $2n+2$ dimension with $n\ge 1$, the higher Chern--Simons action constructed from a semistrict Lie $2$-algebra yields a nonvanishing higher WZW term under higher gauge transformations. The extended Cartan homotopy formula serves as the main computational tool throughout this work. This formalism originates from the anomaly analysis of J. Ma\~nes et al.\cite{algebra}, and was later systematized by F. Izaurieta et al. \cite{seperat} through a subspace separation method that decomposes Chern-Simons forms into gauge--transformed, parameter--dependent, and exact differential pieces. A parallel transgression form approach was developed by P. Mora et al. \cite{tansi}. While these foundational works focused on ordinary gauge theories, Song \cite{danhuaechf} generalized the method to higher Chern--Simons theories based on strict Lie 2-algebras. The present work continues this line of generalization and lifts the extended Cartan homotopy formula to the semistrict Lie 2-algebra setting, where the Jacobiator is generically nonvanishing.

Employing this formalism, we evaluate the variation of the higher Chern--Simons action under arbitrary higher gauge transformations, derive a general decomposition identity, and extract the explicit form of the higher WZW term for general semistrict Lie 2-algebras in arbitrary dimension $2n+2$. We further show that the higher WZW anomaly remains nonvanishing provided its Chevalley--Eilenberg cohomology class is nontrivial. This contrasts with the strict Lie 2-algebra case studied in Ref.~\cite{danhuawzw}, where the vanishing Jacobiator reduces the higher WZW term to vanish identically. Our result thus clarifies that the nonvanishing Jacobiator is the essential ingredient for a nontrivial WZW anomaly in this framework.  
   Our result therefore unifies the two previous observations:
   \begin{itemize}
       \item For strict Lie 2-algebras (Jacobiator $=0$) we recover the result of~\cite{danhuawzw} ( higher WZW term vanishes in any dimension).
       \item For the special case  $2n+2=4$ of semistrict Lie 2-algebras, we recover the explicit computation of~\cite{zucchini4-cs}.
   \end{itemize}
 
   The paper is organised as follows.  
   Sec.~\ref{section 2} briefly recalls the necessary notions of semistrict Lie 2-algebras connections and higher gauge transformation.
   Sec.~ \ref{section 3} introduce the homotopy operator and the construction of higher Chern--Simons actions by Cartan homotopy formula.   
   Sec.~\ref{section 4} contains the main computation: the gauge variation in arbitrary $2n+2$ dimension, leading to the non--vanishing higher WZW term.  
   We conclude in Sec.~\ref{section 5} with a discussion of the implications and possible extensions.
   \section {Semistrict Lie 2-algebra connections and higher gauge transformation}\label{section 2}
  
  We begin by reviewing the fundamental definitions and properties of semistrict Lie 2-algebra, which also called 2-term $L_{\infty}$ algebra $\mathfrak{v}$, then focus on the $\mathfrak{v}$-connections and associated gauge transformations, following the conventions of our previous work~\cite{mengy} and of Refs.~\cite{zucchini4-cs,zucchiniopei,zucchiniopeii}. 
  
  A semistrict Lie 2-algebra  $\mathfrak{v} = (\mathfrak{v}_0,\mathfrak{v}_1,\alpha, [\cdot,\cdot], [\cdot,\cdot], [\cdot,\cdot,\cdot])$ consists of two vector spaces, a map $\alpha:\mathfrak{v}_1\to\mathfrak{v}_0$, a bilinear bracket on $\mathfrak{v}_0$, an action of $\mathfrak{v}_0$ on $\mathfrak{v}_1$, and a trilinear bracket on $\mathfrak{v}_0$ with values in $\mathfrak{v}_1$, satisfying standard coherence identities (see Appendix~\ref{appA}). 
  
  Denote by $\Omega^k(M,\mathfrak{v}_i)$ the space of $\mathfrak{v}_i$-valued $k$-forms. For $A = \sum_a A^a x_a \in \Omega^k(M,\mathfrak{v}_0)$ and $B = \sum_b B^b Y_b \in \Omega^\ell(M,\mathfrak{v}_1)$, we define
  \[
  dA = \sum_a dA^a x_a,\quad 
  [A_1,A_2] = \sum_{ab} A_1^a\wedge A_2^b [x_a,x_b],\quad 
  [A_1,A_2,A_3] = \sum_{abc} A_1^a\wedge A_2^b\wedge A_3^c [x_a,x_b,x_c],
  \]
  \[
  \alpha(B) = \sum_b B^b\,\alpha(Y_b),\quad 
  [A,B] = \sum_{ab} A^a\wedge B^b [x_a,Y_b],
  \]
  with graded symmetry as in~\cite{zucchini4-cs}. 
  
  A $\mathfrak{v}$-connection is a pair $(A,B)$ with $A\in\Omega^1(M,\mathfrak{v}_0)$, $B\in\Omega^2(M,\mathfrak{v}_1)$. Its curvature $(\mathcal{F},\mathcal{H})$ is given by
  \begin{align}
      \mathcal{F} &= dA + \tfrac12[A,A] - \alpha(B), \label{curvF}\\
      \mathcal{H} &= dB + [A,B] - \tfrac16[A,A,A]. \label{curvH}
  \end{align}
  These satisfy the 2-Bianchi identities
  \begin{align}
      d\mathcal{F} + [A,\mathcal{F}] + \alpha(\mathcal{H}) &= 0,\\
      d\mathcal{H} + [A,\mathcal{H}] - [\mathcal{F},B] + \tfrac12[A,A,\mathcal{F}] &= 0.
  \end{align}
  A $\mathfrak{v}$-connection $(A,B)$ is called flat if the corresponding curvature $(\mathcal{F},\mathcal{H})=(0,0)$.
  A finite gauge transformation is parametrized by a set of data $(g,\sigma_g,\Sigma_g,\tau_g)$ subject to the relations detailed in~\cite{mengy}.   A finite semistrict Lie 2-algebra 1-gauge transformation is  orthogonal  if 
   \begin{align}\label{invar}
      \langle g_0(x_1), \cdots ,g_0(x_r); g_1(X) \rangle_{\mathfrak{v}_0 \mathfrak{v}_1}&= \langle x_1,\cdots,x_r, X \rangle_{\mathfrak{v}_0 \mathfrak{v}_1},\\
      \langle  g_0(x_1), \cdots ,g_0(x_r);g_2(y,z) \rangle_{\mathfrak{v}_0 \mathfrak{v}_1}&= -\sum\limits_{i=1}^{r} \langle g_0(x_1), \cdots,g_0(z),\cdots, g_0(x_r) ;g_2(y,x_i) \rangle_{\mathfrak{v}_0 \mathfrak{v}_1},\\
      \langle x_1,\cdots,x_r;\tau_g (y) \rangle_{\mathfrak{v}_0 \mathfrak{v}_1}&=-\sum\limits_{i=1}^{r} \langle x_1, \cdots,y,\cdots, x_r ;\tau_g(x_i) \rangle_{\mathfrak{v}_0 \mathfrak{v}_1}.
  \end{align}
Here \(\langle\,\cdot,\cdot\,\rangle_{\mathfrak{v}_0\mathfrak{v}_1}\) denotes an invariant form (see Appendix \ref{appB}.
  The transformed connection $(A^g,B^g)$ and curvature $(\mathcal{F}^g,\mathcal{H}^g)$ take the compact form
  \begin{align}
      A^g &= g_0\,(A - \sigma_g),\\
      B^g &= g_1\bigl(B - \Sigma_g + \tau_g(A-\sigma_g)\bigr) - \tfrac12 g_2(A-\sigma_g, A-\sigma_g),\\[4pt]
      \mathcal{F}^g &= g_0(\mathcal{F}),\\
      \mathcal{H}^g &= g_1\bigl(\mathcal{H} - \tau_g(\mathcal{F})\bigr) + g_2(A-\sigma_g,\mathcal{F}).
  \end{align}
  Here $g=(g_0,g_1,g_2)$ are the components of an automorphism of $\mathfrak{v}$, $(\sigma_g,\Sigma_g)$ is a flat $\mathfrak{v}$-connection doublet and $\tau_g \in \Omega^1(M,\mathfrak{aut}_1(\mathfrak{v}))$ (see Ref.~\cite{mengy}). In the following sections, we restrict all gauge transformations under consideration to those that are orthogonal.

  \section{ Homotopy operator and semistrict higher Chern-Simons action }\label{section 3}
  In this section, we first introduce the homotopy operator and Cartan homotopy formula. Then, we derive  semistrict higher Chern-Weil theorem and  give $(2n+2)d$ semistrict 2-Chern-Simons form.
  
 Let \((A_{t}, B_{t})\) be the linear interpolation between two $\mathfrak{v}$-connections \((A_{0}, B_{0})\) and \((A_{1}, B_{1})\),
  \begin{align}\label{conn}
      A_{t} &= A_{0}+t \,\eta,\quad \eta:=A_{1}-A_{0},\\
      B_{t} &= B_{0}+t \,\phi,\quad \phi:=B_{1}-B_{0},
  \end{align}
  with \(t\in[0,1]\). 
  It is straightforward to show that\begin{align}
      \frac{\partial  \mathcal{F}_{t}}{\partial t}=d\eta+[A_{t}, \eta]-\alpha(\phi),
  \end{align}
  \begin{align}
      \frac{\partial  \mathcal{H}_{t}}{\partial t}=d\phi+[A_{t},\phi]+[\eta,B_t]-\frac{1}{2}[\eta,A_t,A_t],
  \end{align}
   where $\mathcal{F}_t,\mathcal{H}_t$ are the curvatures of $A_t,B_t$.
 
 The homotopy derivation operator \(l_{t}\) \cite{salwzw,danhuawzw} is defined by
  \begin{equation}
      l_{t} A_{t}=l_{t} B_{t}=0,\quad
      l_{t} \mathcal{F}_{t}=d_{t} A_{t}=dt \frac{\partial}{\partial t} A_{t}=dt\,\eta,\quad
      l_{t} \mathcal{H}_{t}=d_{t} B_{t}=dt \frac{\partial}{\partial t} B_{t}=dt\,\phi.
  \end{equation}
  Then, we extend it as a graded derivation on polynomials in \(A_{t}\), \(B_{t}\), \(\mathcal{F}_{t}\), and \(\mathcal{H}_{t}\). A direct computation gives
  \begin{align}
  (l_{t}d+dl_{t})A_{t}&=l_{t}dA_{t}=l_{t}\bigl(\mathcal{F}_{t}-\frac{1}{2}[A_{t},A_t]+\alpha(B_{t})\bigr)=dt\,\eta , \label{lt:a}\\
      (l_{t} d+d l_{t}) B_{t}&=l_{t} d B_{t}=l_{t}\bigl(\mathcal{H}_{t}-[A_{t}, B_{t}]-\frac{1}{6}[A_t,A_t,A_t]\bigr)=dt\,\phi, \label{lt:b} \\
      (l_{t} d+d l_{t}) \mathcal{F}_{t}
      &=l_{t} d \mathcal{F}_{t}+d l_{t} \mathcal{F}_{t}=l_{t}\bigl([\mathcal{F}_{t}, A_{t}]-\alpha(\mathcal{H}_{t})\bigr)+d l_{t} \mathcal{F}_{t} \notag\\
      &=dt\bigl(d\eta-[\eta,A_t]-\alpha(\phi)\bigr)=dt \frac{\partial}{\partial t} \mathcal{F}_{t}, \label{lt:c}\\
      (l_{t} d +d l_{t}) \mathcal{H}_{t}&=l_{t}\bigl(-[A_t ,\mathcal{H}_t]+[\mathcal{F}_t , B_t]-\frac{1}{2}[A_t,A_t,\mathcal{F}_t]\bigr)+d l_{t} \mathcal{H}_{t} \notag\\
      &=dt\bigl(d\phi+[A_t,\phi]+[\eta,B_t]-\frac{1}{2}[\eta,A_t,A_t]\bigr)=dt \frac{\partial}{\partial t} \mathcal{H}_{t}. \label{lt:d}
      \end{align}
 Consequently, for any polynomial \(\mathcal{P}\) in \(A_{t},B_{t},\mathcal{F}_{t},\mathcal{H}_{t}\), one has
  \begin{equation}\label{eq:cartan-homotopy-S}
      (l_{t}d+dl_{t})\mathcal{P}(A_{t},B_{t},\mathcal{F}_{t},\mathcal{H}_{t})
      =d_{t}\mathcal{P}(A_{t},B_{t},\mathcal{F}_{t},\mathcal{H}_{t})
      =dt\, \frac{\partial}{\partial t}\mathcal{P}(A_{t},B_{t},\mathcal{F}_{t},\mathcal{H}_{t}).
  \end{equation}
  
  Define the homotopy operator \(K_{01}\) as:
  \begin{equation}
      K_{01}:=\int_{0}^{1} l_{t}\,dt.
  \end{equation}
  Integrating \eqref{eq:cartan-homotopy-S} over $t$ from $0$ to $1$ yields the Cartan homotopy formula \cite{danhuaechf}
  \begin{equation}\label{eq:cartan-homotopy}
      \mathcal{P}(A_{1}, B_{1}, \mathcal{F}_{1}, \mathcal{H}_{1})-\mathcal{P}(A_{0}, B_{0}, \mathcal{F}_{0}, \mathcal{H}_{0})
      =\bigl(K_{01} d+d K_{01}\bigr) \mathcal{P}(A_{t}, B_{t}, \mathcal{F}_{t}, \mathcal{H}_{t}).
  \end{equation}
  
  We now choose the invariant polynomial
  \begin{equation}
      \mathcal{P}_{2 n+3}\bigl(\mathcal{F}_{t}, \mathcal{H}_{t}\bigr):=\bigl\langle\mathcal{F}_{t}^{n}, \mathcal{H}_{t}\bigr\rangle_{\mathfrak{v}_0\mathfrak{v}_1},
  \end{equation}
  where $\mathcal{F}_t,\mathcal{H}_t$ are the curvatures of $A_t,B_t$ in \eqref{conn} and \(\langle\,\cdot,\cdot\,\rangle_{\mathfrak{v}_0\mathfrak{v}_1}\) denotes an invariant form (see Appendix \ref{appB} ) . This form is closed:
  \begin{equation}
      d\bigl\langle\mathcal{F}_{t}^{n}, \mathcal{H}_{t}\bigr\rangle_{\mathfrak{v}_0\mathfrak{v}_1}=0.
  \end{equation}
  Substituting \(\mathcal{P}=\mathcal{P}_{2 n+3}\) into \eqref{eq:cartan-homotopy}, we obtain the semistrict higher Chern-Weil theorem
  \begin{equation}\label{eq:higher-chern-weil}
      \bigl\langle\mathcal{F}_{1}^{n}, \mathcal{H}_{1}\bigr\rangle_{\mathfrak{v}_0\mathfrak{v}_1}
      -\bigl\langle\mathcal{F}_{0}^{n}, \mathcal{H}_{0}\bigr\rangle_{\mathfrak{v}_0\mathfrak{v}_1}
      =d\mathcal{Q}_{2 n+2}(A_{1}, B_{1} ; A_{0}, B_{0}).
  \end{equation}
  Here, the $(2n+2)$-form $\mathcal{Q}_{2n+2}(A_{1},B_{1};A_{0},B_{0})$ is given by
  \begin{align}\label{eq:higher-transgression}
      \mathcal{Q}_{2n+2}(A_{1},B_{1};A_{0},B_{0})
      &=K_{01} \langle\mathcal{F}_{t}^{n}, \mathcal{H}_{t}\rangle_{\mathfrak{v}_0\mathfrak{v}_1}\\
      &=\int\limits_{0}^{1} dt\,\bigg( n \langle A_1-A_0, \mathcal{F}_{t}^{n-1};\mathcal{H}_{t} \rangle_{\mathfrak{v}_0 \mathfrak{v}_1}
      + \langle \mathcal{F}_{t}^{n};B_1-B_0 \rangle_{\mathfrak{v}_0 \mathfrak{v}_1}\bigg),
  \end{align}
  where we have used the basic property of $l_t$ given in Eq.~\eqref{lt:a},\eqref{lt:b},\eqref{lt:c},\eqref{lt:d}.
  
 Setting \(A_{0}=B_{0}=0\), \(A_{1}=A\), \(B_{1}=B\). A direct calculation gives the \textbf{ $\bm{(2n+2)d}$ semistrict 2-Chern-Simons form}
 \begin{align}\label{2cs}
     \mathcal{CS}_{2n+2}(A,B,\mathcal{F},\mathcal{H})
     &=\int\limits_{0}^{1} dt( n \langle A, \mathcal{F}_{t}^{n-1};\mathcal{H}_{t} \rangle_{\mathfrak{v}_0 \mathfrak{v}_1}+
     \langle \mathcal{F}_{t}^{n};B \rangle_{\mathfrak{v}_0 \mathfrak{v}_1})\\\nonumber
     &=\frac{n}{2^{n}}\langle A,(dA+\frac{1}{3}[A,A]-\alpha(B) )^{n-1};dB+\frac{2}{3}[A,B]-\frac{1}{12}[A,A,A] \rangle_{\mathfrak{v}_0 \mathfrak{v}_1} \\\nonumber
     &\quad+\frac{1}{2^{n}}\langle  (dA+\frac{1}{3}[A,A]-\alpha(B))^{n};B \rangle_{\mathfrak{v}_0 \mathfrak{v}_1}.
 \end{align}
  
  \section{Semistrict higher Wess–Zumino–Witten terms}\label{section 4}
  In this section, we study the transformation of the semistrict higher CS form under higher gauge transformations and derive the semistrict higher WZW term.
  
  Under a higher gauge transformation specified by $g=(g_0,g_1,g_2),\sigma_g,\Sigma_g,\tau_g$, the 2-connection transforms as
  \begin{align}
      &A^{g}=g_0(A-\sigma_g),\\\nonumber
      &B^{g}=g_1(B-\Sigma_g+\tau_{g}(A-\sigma_g))-\frac{1}{2}g_2(A-\sigma_g,A-\sigma_g).
  \end{align}
 Denote
  \begin{equation}
 V:=-\sigma_g\quad
 W:=-\Sigma_g+\tau_g(V)-\frac{1}{2}g_1^{-1}g_2(V,V).
  \end{equation}
  The corresponding curvatures transform as
  \begin{align}
     &\mathcal{F}^{g}=g_0(\mathcal{F}),\\\nonumber
     &\mathcal{H}^{g}=g_1(\mathcal{H}-\tau_{g}(\mathcal{F}))+g_2(A-\sigma_g,\mathcal{F}).
      \end{align}
Let $A_0=g_0(V),B_0=g_1(W),A_1=A^g,B_1=B^g$, we have
  \begin{align}
      A_{t}^{g}&=A_0+t(A_1-A_0)=g_0(A_t-\sigma_g),\label{eq:homotopy-A}\\
      B_{t}^{g}&=B_0+t(B_1-B_0)\\
      &=g_1(B_t+W+\tau_{g}(A_t))-\frac{1}{2}g_2(A_t,A_t)+g_2(A_t,V),\label{eq:homotopy-B}
  \end{align}
  with \(A_{t}=t A\), \(B_{t}=t B\), \(t\in[0,1]\).
   We obtain
  \begin{align}
     & \mathcal{F}_{t}^{g}=g_0(\mathcal{F}_t),\\
     &\mathcal{H}_t^{g}=g_1(\mathcal{H}_t-\tau_{g}(\mathcal{F}_t))+g_2(A_t-\sigma_g,\mathcal{F}_t),
    \end{align}
 where $\mathcal{F}_{t},\mathcal{H}_t$ are the corresponding curvature of $A_t,B_t$.

   Substituting $\mathcal{P}=\mathcal{CS}_{2n+2}(A_{t}^{g},B_{t}^{g},\mathcal{F}_{t}^{g},\mathcal{H}_{t}^{g})$ into \eqref{eq:cartan-homotopy}, we obtain
 \begin{align}\label{eq:CS-variation-main}
        &\mathcal{CS}_{2 n+2}\bigl(A^{g}, B^{g}, \mathcal{F}^{g}, \mathcal{H}^{g}\bigr)
         -\mathcal{CS}_{2 n+2}\bigl(g_0(V), g_1 (W), 0,0\bigr)\\
             &=\bigl(K_{01} d+d K_{01}\bigr) \mathcal{CS}_{2 n+2}\bigl(A_{t}^{g}, B_{t}^{g}, \mathcal{F}_{t}^{g}, \mathcal{H}_{t}^{g}\bigr)
      \end{align}
Suppose $A_0=B_0=0,A_1=A^g,B_1=B^g$, we write
   \begin{align}
          &\mathcal{CS}_{2n+2}\bigl(A^{g},B^{g},\mathcal{F}^{g},\mathcal{H}^{g}\bigr)\\
          &=\int _{0}^{1}ds\biggl\{ n\bigl\langle A^{g}\wedge (\mathcal{F}_{s}^{g})^{n-1},\mathcal{H}_{s}^{g}\bigr\rangle_{\mathfrak{v}_0\mathfrak{v}_1}
          +\bigl\langle(\mathcal{F}_{s}^{g})^{n},B^{g}\bigr\rangle_{\mathfrak{v}_0\mathfrak{v}_1}\biggr\},
      \end{align}
  where $\mathcal{F}_s ,\mathcal{H}_s$ are the curvature of $sA^g,sB^g$.
Similarly, setting $A_0=B_0=0,A_1=A_t^g,B_1=B_t^g$, we have
   \begin{align}
      &\mathcal{CS}_{2n+2}\bigl(A_t^{g},B_t^{g},\mathcal{F}_t^{g},\mathcal{H}_t^{g}\bigr)\\
      &\quad=\int _{0}^{1}ds\biggl\{ n\bigl\langle A_t^{g}\wedge (\mathcal{F}_{ts}^{g})^{n-1},\mathcal{H}_{ts}^{g}\bigr\rangle_{\mathfrak{v}_0\mathfrak{v}_1}
      +\bigl\langle(\mathcal{F}_{ts}^{g})^{n},B_t^{g}\bigr\rangle_{\mathfrak{v}_0\mathfrak{v}_1}\biggr\},
  \end{align}
  where $\mathcal{F}_{ts} ,\mathcal{H}_{ts}$ are the curvature of $sA_{t}^g,sB_{t}^g$.
  
  For the right-hand side of \eqref{eq:CS-variation-main}, we have verified that
  \begin{equation}
      d\mathcal{CS}_{2n+2}\bigl(A_{t}^{g},B_{t}^{g},\mathcal{F}_{t}^{g},\mathcal{H}_{t}^{g}\bigr)
      =\bigl\langle(\mathcal{F}_{t}^{g})^{n},\mathcal{H}_{t}^{g}\bigr\rangle_{\mathfrak{v}_0\mathfrak{v}_1}
      =\bigl\langle(\mathcal{F}_{t})^{n},\mathcal{H}_{t}\bigr\rangle_{\mathfrak{v}_0\mathfrak{v}_1}.
  \end{equation}
  Consequently,
  \begin{equation}
      \begin{aligned}
          K_{01} d \mathcal{CS}_{2 n+2}\bigl(A_{t}^{g}, B_{t}^{g}, \mathcal{F}_{t}^{g}, \mathcal{H}_{t}^{g}\bigr)
          &=K_{01}\bigl\langle (\mathcal{F}_{t})^{n}, \mathcal{H}_{t}\bigr\rangle_{\mathfrak{v}_0\mathfrak{v}_1}
          =\int_{0}^{1} l_{t}\bigl\langle (\mathcal{F}_{t})^{n}, \mathcal{H}_{t}\bigr\rangle_{\mathfrak{v}_0\mathfrak{v}_1}dt\\
          &=\mathcal{CS}_{2n+2}(A, B, \mathcal{F}, \mathcal{H})=\mathcal{CS}_{2n+2}(A,B).
      \end{aligned}
  \end{equation}
 We have thus identified the \((2n+1)\)-form $\beta_{2n+1}$:
  \begin{equation}
      \beta_{2n+1}:=K_{01}\mathcal{CS}_{2n+2}\bigl(A_{t}^{g},B_{t}^{g},\mathcal{F}_{t}^{g},\mathcal{H}_{t}^{g}\bigr)
      =\int_{0}^{1}l_{t}\mathcal{CS}_{2n+2}\bigl(A_{t}^{g},B_{t}^{g},\mathcal{F}_{t}^{g},\mathcal{H}_{t}^{g}\bigr)dt.
  \end{equation}
  Then, the Cartan homotopy formula simplifies to
  \begin{align}\label{eq:CS-variation-simplified}
          \mathcal{CS}_{2 n+2}(A^{g}, B^{g})
          -\mathcal{CS}_{2n+2}(A,B)=\mathcal{CS}_{2n+2}(g_0(V),g_1(W))+d\beta_{2n+1},
   \end{align}
where
\begin{align}
    \beta_{2n+1}
    &=\int_{0}^{1} l_t \big\{\frac{n}{2^{n}}\langle A_t^g,(dA_t^g+\frac{1}{3}[A_t^g,A_t^g]-\alpha(B_t^g) )^{n-1};dB_t^g+\frac{2}{3}[A_t^g,B_t^g]-\frac{1}{12}[A_t^g,A_t^g,A_t^g] \rangle_{\mathfrak{v}_0 \mathfrak{v}_1} \\\nonumber
    &\quad+\frac{1}{2^{n}}\langle  (dA_t^g+\frac{1}{3}[A_t^g,A_t^g]-\alpha(B_t^g))^{n};B_t^g \rangle_{\mathfrak{v}_0 \mathfrak{v}_1}\}\\\nonumber
    &=\int_{0}^{1} l_t \big\{\frac{n}{2^{n}}\langle A_t^g,(\mathcal{F}_t^g-\frac{1}{6}[A_t^g,A_t^g] )^{n-1};\mathcal{H}_t^g-\frac{1}{3}[A_t^g,B_t^g]+\frac{1}{12}[A_t^g,A_t^g,A_t^g] \rangle_{\mathfrak{v}_0 \mathfrak{v}_1} \\\nonumber
    &\quad+\frac{1}{2^{n}}\langle  (\mathcal{F}_t^g-\frac{1}{6}[A_t^g,A_t^g])^{n};B_t^g \rangle_{\mathfrak{v}_0 \mathfrak{v}_1}\}\\\nonumber
    &=\int_{0}^{1} dt \big\{\frac{n(n-1)}{2^{n}}\langle A_t^g,g_0(A),(\mathcal{F}_t^g-\frac{1}{6}[A_t^g,A_t^g])^{n-2} ;\mathcal{H}_t^g-\frac{1}{3}[A_t^g,B_t^g]+\frac{1}{12}[A_t^g,A_t^g,A_t^g]\rangle_{\mathfrak{v}_0 \mathfrak{v}_1}\\\nonumber
    &+\frac{n}{2^{n}}\langle A_t^g,(\mathcal{F}_t^g-\frac{1}{6}[A_t^g,A_t^g] )^{n-1};g_1(B+\tau_g(A))-tg_2(A,A)+g_2(A,\sigma_g) \rangle_{\mathfrak{v}_0 \mathfrak{v}_1} \\\nonumber
    &+\frac{1}{2^{n}}\langle  g_0(A),(\mathcal{F}_t^g-\frac{1}{6}[A_t^g,A_t^g])^{n-1};B_t^g \rangle_{\mathfrak{v}_0 \mathfrak{v}_1}\}.
\end{align}

In particular,  we consider $A=\sigma_g,B=\Sigma_g$. Then, $A^g=0,B^g=0$ and Eq. \eqref{eq:CS-variation-simplified} becomes
 \begin{align}\label{2}
  -\mathcal{CS}_{2n+2}(\sigma_g,\Sigma_g)=\mathcal{CS}_{2n+2}(g_0(V),g_1(W))+d\beta'_{2n+1}
\end{align} 
with
\begin{align}
    \beta'_{2n+1}
    &=\int_{0}^{1} dt \big\{\frac{n(n-1)}{2^{n}}\langle (t-1)g_0(\sigma_g),g_0(\sigma_g),(-\frac{(t-1)^2}{6}[g_0(\sigma_g),g_0(\sigma_g)])^{n-2};\\\nonumber &-\frac{(t-1)^2}{3}[g_0(\sigma_g),g_1(\Sigma_g+\tau_g(\sigma_g))-\frac{t-1}{2}g_2(\sigma_g,\sigma_g)]+\frac{(t-1)^3}{12}[g_0(\sigma_g),g_0(\sigma_g),g_0(\sigma_g)]\rangle_{\mathfrak{v}_0 \mathfrak{v}_1}\\\nonumber
    &+\frac{n}{2^{n}}\langle (t-1)g_0(\sigma_g),(-\frac{(t-1)^2}{6}[g_0(\sigma_g),g_0(\sigma_g)])^{n-1};g_1(\Sigma_g+\tau_g(\sigma_g))-(t-1)g_2(\sigma_g,\sigma_g) \rangle_{\mathfrak{v}_0 \mathfrak{v}_1} \\\nonumber
    &+\frac{(t-1)^2}{2^{n}}\langle  g_0(\sigma_g),(-\frac{(t-1)^2}{6}[g_0(\sigma_g),g_0(\sigma_g)])^{n-1};g_1(\Sigma_g+\tau_g(\sigma_g))-\frac{t-1}{2}g_2(\sigma_g,\sigma_g) \rangle_{\mathfrak{v}_0 \mathfrak{v}_1}\}\\\nonumber% 
    &=\frac{n(n-1)}{2^{n}}\langle g_0(\sigma_g),g_0(\sigma_g),([g_0(\sigma_g),g_0(\sigma_g)])^{n-2};\frac{(-1)^{n}}{6^{n-1}n}[g_0(\sigma_g),g_1(\Sigma_g+\tau_g(\sigma_g))]\\\nonumber
    &+\frac{(-1)^{n}}{6^{n-1}(2n)}[g_0(\sigma_g),g_2(\sigma_g,\sigma_g)]+\frac{(-1)^{n}}{6^{n-1}(2n+1)2}[g_0(\sigma_g),g_0(\sigma_g),g_0(\sigma_g)]\rangle_{\mathfrak{v}_0 \mathfrak{v}_1}\\\nonumber
    &+\frac{n}{2^{n}}\langle g_0(\sigma_g),([g_0(\sigma_g),g_0(\sigma_g)])^{n-1};\frac{(-1)^{n}}{6^{n-1}(2n)}g_1(\Sigma_g+\tau_g(\sigma_g))+\frac{(-1)^{n}}{6^{n-1}(2n+1)}(\sigma_g,\sigma_g) \rangle_{\mathfrak{v}_0 \mathfrak{v}_1} \\\nonumber
    &+\frac{1}{2^{n}}\langle  g_0(\sigma_g),[g_0(\sigma_g),g_0(\sigma_g)])^{n-1};\frac{(-1)^{n-1}}{6^{n-1}(2n+1)}g_1(\Sigma_g+\tau_g(\sigma_g))+\frac{(-1)^{n-1}}{6^{n-1}2(2n+2)}g_2(\sigma_g,\sigma_g) \rangle_{\mathfrak{v}_0 \mathfrak{v}_1}  .
\end{align}
Substituting Eq.~\eqref{2} into Eq.~\eqref{eq:CS-variation-simplified}, we obtain
\begin{align}
    \mathcal{CS}_{2n+2}(A^g,B^g) - \mathcal{CS}_{2n+2}(A,B) 
    &= -\mathcal{CS}_{2n+2}(\sigma_g,\Sigma_g) + d\gamma_{2n+1},
\end{align}
where $\gamma_{2n+1} = \beta_{2n+1} - \beta'_{2n+1}$. From Eq.\eqref{2cs}, we derive the explicit expression for $\mathcal{CS}_{2n+2}(\sigma_g,\Sigma_g)$.
\begin{align}
    &\mathcal{CS}_{2n+2}(\sigma_g,\Sigma_g)\nonumber\\
    &=\frac{n}{2^{n}}\langle \sigma_g,(-\frac{1}{6}[\sigma_g,\sigma_g] )^{n-1};-\frac{1}{3}[\sigma_g,\Sigma_g]+\frac{1}{12}[\sigma_g,\sigma_g,\sigma_g] \rangle_{\mathfrak{v}_0 \mathfrak{v}_1}+\frac{1}{2^{n}}\langle  (-\frac{1}{6}[\sigma_g,\sigma_g])^{n};\Sigma_g \rangle_{\mathfrak{v}_0 \mathfrak{v}_1}\\\nonumber
    &=\frac{(-1)^n n }{ 2^{n+1} 9}\langle ([\sigma_g,\sigma_g] )^{n};\Sigma_g \rangle_{\mathfrak{v}_0 \mathfrak{v}_1}
   +\langle \frac{(-1)^{n-1} n }{2^{n+3} 9}  \sigma_g+\frac{(-1)^{n-1} n(n-1) }{ 2^{n+1}27} \alpha(\Sigma_g), ([\sigma_g,\sigma_g])^{n-1};[\sigma_g,\sigma_g, \sigma_g] \rangle_{\mathfrak{v}_0 \mathfrak{v}_1}.
\end{align}

The higher WZW term $\mathcal{CS}_{2n+2}(\sigma_g,\Sigma_g)$ is closed and $(\sigma_g,\Sigma_g)$ is a flat connection doublet for any $g$. With $\mathcal{CS}_{2n+2}(\sigma_g,\Sigma_g)$ there is therefore associated a special higher Chevalley--Eilenberg cochain $\chi \in \mathrm{CE}^{2n+2}(\mathfrak{v})$,
    \begin{align}
    \chi&=\frac{n}{2^{n}}\langle \pi,(-\frac{1}{6}[\pi,\pi] )^{n-1};-\frac{1}{3}[\pi,\Pi]+\frac{1}{12}[\pi,\pi,\pi] \rangle_{\mathfrak{v}_0 \mathfrak{v}_1} +\frac{1}{2^{n}}\langle  (-\frac{1}{6}[\pi,\pi])^{n};\Pi \rangle_{\mathfrak{v}_0 \mathfrak{v}_1}.
\end{align}
    which is in fact a cocycle.  if $\chi$ is exact in $\mathrm{CE}(\mathfrak{v})$, then $\mathcal{CS}_{2n+2}(\sigma_g,\Sigma_g)$ is exact in $\Omega_{\mathfrak{v}}^*(M)$. In this way, in order the anomaly $\mathcal{CS}_{2n+2}(\sigma_g,\Sigma_g)$ to be non trivial, it is necessary that $H_{\mathrm{CE}}^{2n+2}(\mathfrak{v}) \neq 0$. 
Just as the WZW term arising from a gauge transformation in ordinary three dimensional Chern–Simons theory is precisely the Chern–Simons action of the flat connection $g^{-1}dg$ associated with the gauge parameter $g$ \cite{ordwzw}, in our semistrict higher gauge theory the \textbf{higher Wess–Zumino–Witten term} is exactly the topological quantity $-\mathcal{CS}_{2n+2}(\sigma_g,\Sigma_g)$, corresponding to the higher Chern–Simons action of the flat $\mathfrak{v}$-connection $(\sigma_g,\Sigma_g)$ for the higher gauge parameter.

  Explicitly for $n=1$, we have 
  \begin{align}
      \mathcal{CS}_{4}((A,B))&=\frac{1}{2}\langle 2\mathcal{F}+\alpha(B);B \rangle_{\mathfrak{v}_0 \mathfrak{v}_1}-\frac{1}{24}\langle A;[A,A,A] \rangle_{\mathfrak{v}_0 \mathfrak{v}_1}-d\langle A,B \rangle_{\mathfrak{v}_0 \mathfrak{v}_1},\\
      \beta_{2n+1}&=\frac{1}{2} \langle 2g_0(A), B_{t}^{g} \rangle_{\mathfrak{v}_0\mathfrak{v}_1},\quad
      \beta'_{2n+1}=\frac{1}{2} \langle 2g_0(\sigma_g), B_{t}^{\prime g} \rangle_{\mathfrak{v}_0\mathfrak{v}_1}.
  \end{align}
   where \begin{align}
     B_t^g&=\int_{0}^1 dt [g_1(tB-\Sigma_g+\tau_{g}(tA-\sigma_g))-\frac{1}{2}g_2(tA-\sigma_g,tA-\sigma_g)],\\
      B_t^{\prime g}&= \int_{0}^1 dt [(t-1)g_1(\Sigma_g+\tau_{g}(\sigma_g))-\frac{1}{2}(t-1)^2 g_2(\sigma_g,\sigma_g)].
   \end{align}
   Then, we have 
   \begin{align}\label{exact}
       \gamma_{2n+1}=\frac{1}{6} \langle A- \sigma_g, g_1^{-1} g_2(A - \sigma_g, A - \sigma_g) + 6\Sigma_g - 3\tau_g(A-\sigma_g)\rangle_{\mathfrak{v}_0\mathfrak{v}_1}.
   \end{align}
This result is the same as Ref.\cite{zucchini4-cs}.
In particular, when the three bracket is set to zero (i.e., the algebra reduces to the strict crossed module) and the gauge transformation is appropriately restricted (see Appendix~\ref{appA}), our result consistently reduces to that of Ref.~\cite{danhuawzw} for the strict higher-dimensional case, up to group inversion, sign conventions, and relabeling of gauge parameters. Furthermore, in the four-dimensional strict case $(n=1)$, expression \eqref{exact} reproduces the result in Ref.~\cite{schenkel5d}.

\section{Conclusion and outlook}\label{section 5}
In this paper, we construct the semistrict higher Chern--Simons form though the higher Cartan homotopy formula, and derive the explicit expression of the semistrict higher Wess--Zumino--Witten (WZW) term under finite higher gauge transformations. By restricting the three bracket structure, our results naturally reduce to those obtained in the crossed module, which are consistent with the existing literature \cite{danhuawzw}. In particular, the nontrivial contribution of the higher WZW term emerging in the semistrict case reveals distinct topological features that are absent in the strict counterpart, which constitutes the key innovation of the present work.

While the present discussion is restricted to Lie 2-algebras, a systematic treatment of finite gauge transformations and their associated anomalies for general $L_{\infty}$ algebras remains to be further developed. It would be interesting to extend the current construction to general $n$-term $L_{\infty}$ algebras, exploring their invariant polynomial structures, and investigating the corresponding chiral anomalies and topological effective actions will be left for future investigation. Likewise, applying our coherence framework to physically motivated models appears to be a natural next step.

\begin{appendices}
  
    \section{Semistrict Lie 2-algebra}\label{appA}

    The definitions collected in this appendix are standard in the literature (see e.g.~\cite{zucchini4-cs}) and are also presented in our previous work~\cite{mengy}. They are repeated here for the convenience of the reader.

    A semistrict Lie 2-algebra $\mathfrak{v}$ consists of two real vector spaces $\mathfrak{v}_0$ and $\mathfrak{v}_1$ together with the following linear maps:
    \begin{itemize}
        \item  $\alpha:\mathfrak{v}_1 \rightarrow \mathfrak{v}_0$
        \item  $[\cdot, \cdot]\colon \mathfrak{v}_0 \wedge \mathfrak{v}_0 \to \mathfrak{v}_0$
        \item $[\cdot, \cdot]\colon \mathfrak{v}_0 \otimes \mathfrak{v}_1 \to \mathfrak{v}_1$
        \item $[\cdot, \cdot, \cdot]\colon \mathfrak{v}_0 \wedge \mathfrak{v}_0 \wedge \mathfrak{v}_0 \to \mathfrak{v}_1$
    \end{itemize}
    which are required to satisfy the following relations 
    \begin{align}
        &\alpha([x,X])-[x,\alpha(X)]=0,\\
        &[\alpha (X),Y]+[\alpha (Y),X]=0,\\
        &[x,[y,z]]+[y,[z,x]]+[z,[x,y]]-\alpha([x,y,z])=0,\\
        &[x,[y,X]]-[y,[x,X]]-[[x,y],X]-[x,y, \alpha (X)]=0,\\
        &[x,y,[z,t]]+[x,z,[t,y]]+[x,t,[y,z]]-[y,z,[t,x]]-[z,t,[y,x]]-[t,y,[z,x]]\\\nonumber
        &-[x,[y,z,t]]+[y,[z,t,x]]-[z,[t,x,y]]+[t,[x,y,z]]=0.    
    \end{align}
    for all $x,y,z,t\in \mathfrak{v}_0,X,Y \in \mathfrak{v}_1.$
 
 A finite semistrict Lie 2-algebra 1-gauge transformation consists of the following set of data.
    \begin{itemize}
        \item a map $g=(g_0,g_1,g_2) \in \text{Map}(M,\text{Aut}_1(\mathfrak{v}))$, where $\text{Aut}_1(\mathfrak{v})$ is the set of all automorphism of $\mathfrak{v}$.
        
        \item a flat connection doublet $(\sigma_g,\Sigma_g)$, i.e.,
        \begin{align}
            &d\sigma_g+\frac{1}{2}[\sigma_g , \sigma_g]-\alpha(\Sigma_g)=0,\\
            &d\Sigma_g+[\sigma_g , \Sigma_g]-\frac{1}{6}[\sigma_g , \sigma_g,\sigma_g]=0.
        \end{align}
        \item an element $\tau_g \in \Omega^1(M,\mathfrak{aut}_1(\mathfrak{v}))$, where $\mathfrak{aut}_1(\mathfrak{v})$ is the set of all 2-derivations of $\mathfrak{v}$ satisfying
        \begin{align}
            d\tau_g(x)+[\sigma_g,\tau_g(x)]-[x,\Sigma_g]+\frac{1}{2}[\sigma_g,\sigma_g,x]+\tau_g([\sigma_g,x]+\alpha(\tau_g(x)))=0.
        \end{align}
    \end{itemize}
    These data are required to satisfy the following relations
    \begin{align}
        &g_0^{-1}dg_0(x)-[\sigma_g,x]-\alpha(\tau_g(x))=0,\\
        &g_1^{-1}dg_1(X)-[\sigma_g,X]-\tau_g(\alpha(X))=0,\\
        &g_1^{-1}(dg_2(x,y)-2g_2(g_0^{-1} dg_0(x),y))
        -[\sigma_g,x,y]-\tau_g([x,y])+[x,\tau_g(y)]-[y,\tau_g(x)]=0.
    \end{align}

A differential Lie crossed module consists of the following elements.
    \begin{itemize}
        \item A pair of Lie algebras $\mathfrak{g},\mathfrak{h}$,  
        \item A Lie algebra morphism $t :\mathfrak{h} \rightarrow \mathfrak{g}$,
        \item A Lie algebra morphism $\mu : \mathfrak{g}\rightarrow \text{Der}(\mathfrak{h})$, where Der($\mathfrak{h}$) is the Lie algebra of derivations of $\mathfrak{h}$.
    \end{itemize}
    These are required to satisfy the following relations:
    \begin{align}
        &t(\mu(g)(h))=[g,t(h)],\\
        &\mu(t(h)(h'))=[h,h'].
    \end{align}
    
    A semistrict Lie 2-algebra $\mathfrak{v}$ is strict if $[\cdot, \cdot, \cdot]=0$. There exists a one-to-one correspondence between strict 2-term $L_{\infty}$
    and differential Lie crossed modules.
    \begin{itemize}
        \item $\mathfrak{v}_{0}=\mathfrak{g}$,
        \item $\mathfrak{v}_{1}=\mathfrak{h}$, 
        \item $\alpha(X)=t(X)$,
        \item  $[x,y]=[x,y]_{\mathfrak{g}}$, 
        \item  $[x,X]=\mu(x)(X)$,
        \item $[x,y,z]=0$.
    \end{itemize}

    A crossed module 1-gauge transformation contains the following data:
    \begin{itemize}
        \item A map $\gamma \in Map(M,G)$,  
        \item An element $\phi \in \Omega^{1} (M,\mathfrak{h})$.
    \end{itemize}
    It acts on the connection doublet $(A,B)$ as
    \begin{align}
        A^{\gamma}&=\gamma A\gamma^{-1}-d\gamma\gamma^{-1}-t (\phi ),\\
        B^{\gamma}&=\mu(\gamma)(B)-\mu(A^{\gamma})(\phi)-d\phi-\frac{1}{2}[\phi,\phi].
    \end{align}
    Under this transformation the curvature forms change as  
    \begin{align}
        \mathcal{F}^{\gamma}&=\gamma \mathcal{F} \gamma^{-1},\\
        \mathcal{H}^{\gamma}&=\mu(\gamma)(\mathcal{H})-\mu(\mathcal{F}^{\gamma})(\phi).
    \end{align}
    As noted above, the differential crossed modules correspond to the strict 2-term $L_{\infty}$ algebras; analogously, the crossed module 1-gauge transformation can be obtained by specifying a concrete form of 
    \begin{align}
        g_0&=ad_\gamma, \quad g_1=\mu(\gamma)(\cdot),\quad g_2=0,\\
        \sigma_g&=\gamma^{-1}d\gamma+\gamma^{-1}t(\phi)\gamma,\\
        \Sigma_g&=\mu(\gamma^{-1})(d\phi+\frac{1}{2}[\phi,\phi]),\\
        \tau_g(x)&=\mu(x)(\mu(\gamma^{-1})(\phi)).
    \end{align} 
    
   \section{Symmetric invariant form} \label{appB}
    The brackets $\langle\cdots;\cdot\rangle_{\mathfrak{v}_0\mathfrak{v}_1}$ stand for a balanced multilinear symmetric form for balanced semistrict Lie 2-algebra $\mathfrak{v}$ (i.e. dim $\mathfrak{v}_0$ = dim $\mathfrak{v}_1$)
    \begin{align*}
        \langle\cdots;\cdot\rangle_{\mathfrak{v}_0\mathfrak{v}_1}:\underbrace{\mathfrak{v}_0 \otimes \cdots \otimes \mathfrak{v}_0}_{r} \otimes \mathfrak{v}_1 \rightarrow \mathbb{R}
    \end{align*}
    satisfying
    \begin{align}
        \langle x_1,\cdots,x_i,\cdots,x_r;[x, X] \rangle_{\mathfrak{v}_0\mathfrak{v}_1}&=-\sum\limits_{i=1}^{r}\langle x_1,\cdots,[x,x_i],\cdots,x_r; X\rangle_{\mathfrak{v}_0\mathfrak{v}_1},\label{A.1}\\
        \langle x_1,\cdots,\alpha(Y_i),\cdots,x_r; Y\rangle_{\mathfrak{v}_0\mathfrak{v}_1}&=\langle x_1,\cdots,\alpha(Y),\cdots,x_r; Y_i\rangle _{\mathfrak{v}_0\mathfrak{v}_1}, \label{A.2}\\
        \langle x_1,\cdots,x_i,\cdots,x_r;[x,y,z] \rangle_{\mathfrak{v}_0\mathfrak{v}_1}&=-\sum\limits_{i=1}^{r} \langle x_1,\cdots,z,\cdots,x_r; [x,y,x_i]\rangle_{\mathfrak{v}_0\mathfrak{v}_1} \label{A.3}.
    \end{align}
    $\langle\cdots,\cdot\rangle_{\mathfrak{v}_0\mathfrak{v}_1}$ is symmetric if
    \begin{align}\label{A.4}
        \langle x_1,\cdots,x_i,\cdots,x_j,\cdots,x_r; Y \rangle_{\mathfrak{v}_0\mathfrak{v}_1}= \langle x_1,\cdots,x_j,\cdots,x_i,\cdots,x_r; Y \rangle_{\mathfrak{v}_0\mathfrak{v}_1}.
    \end{align}

\end{appendices}


\begin{thebibliography}{30}
    
    \bibitem{WessZumino1971} \href{https://doi.org/10.1016/0370-2693(71)90582-X}{J. Wess, B. Zumino, Phys. Lett. B 37, 95 (1971).}
   
   
   \bibitem{witten1983global} \href{https://doi.org/10.1016/0550-3213(83)90063-9}{E. Witten, Global aspects of current algebra, Nucl. Phys. B 223 (1983) 422-432.}
    
\bibitem{ordwzw} B. Zumino, Chiral anomalies and differential geometry,
in: B.S. DeWitt, R. Stora (Eds.),
Relativity, Groups and Topology II, Les Houches Summer School 1983, North-Holland, Amsterdam, 1984.

 \bibitem{baze2004higherv} \href{https://doi.org/10.48550/arXiv.math/0307200}{J. Baez , A. Lauda, Theory Appl. Categories 12, (2004) 423.} 

\bibitem{baze2004highervi} \href{http://eudml.org/doc/124217}{J. C. Baez, A. S. Crans,  Theory Appl. Categories 12, (2004) 492.} 

 \bibitem{lada1993sh} \href{http://doi.org/ 10.1007/BF00671791}{T. Lada, J. Stasheff,  Int. J. Theor. Phys. 32, (1993) 1087.} 


\bibitem{lada1995strongly} \href{http://doi.org/10.1080/00927879508825335}{T. Lada, M. Markl, Commun. Algebra 23, (1995) 2147.} 

  \bibitem{danhua2023} \href{https://doi.org/10.1007/JHEP07(2023)207}{D. H. Song, M. Y. Wu, K. Wu, J. Yang, JHEP 07, (2023) 207.} 
 
 \bibitem{danhua2024higher} \href{https://doi.org/10.1016/j.physletb.2023.138374}{D. H. Song, K. Wu, J. Yang, Phys. Lett. B 848, (2024) 138374.} 
 
\bibitem{zucchini4-cs} \href{https://doi.org/10.1063/1.4947531}{R. Zucchini, J. Math. Phys. 57, (2016) 052301. }

\bibitem{zucchiniopei} \href{http://doi.org/10.1016/j.geomphys.2020.103826}{R. Zucchini, J. Geom. Phys. 156, (2020) 103826.}

\bibitem{zucchiniopeii} \href{https://doi.org/10.1016/j.geomphys.2020.103825}{R. Zucchini, J. Geom. Phys. 156, (2020) 103825.}

\bibitem{zucchiniaksz} \href{https://doi.org/10.1007/JHEP03(2013)014}{R. Zucchini, JHEP. 03,(2013) 014.}

\bibitem{zucchiniwilson} \href{ https://doi.org/10.48550/arXiv.1903.02853}{R. Zucchini, arxiv. 1903.02853}.


\bibitem{schenkel5d} \href{ https://doi.org/10.1007/s00220-024-05170-9}{A. Schenkel, B. Vicedo, Commun. Math. Phys. 405, (2024) 293. }
\bibitem{danhuawzw} \href{10.1016/j.physletb.2026.140568}{D. H. Song, Phys. Lett. B 878 (2026) 140568}

 \bibitem{soncini2014} \href{https://doi.org/10.1007/JHEP10(2014)079}{E. Soncini, R. Zucchini, JHEP. 10, (2014) 079.}

\bibitem{salwzw} \href{https://doi.org/10.1016/j.physletb.2013.11.023}{P. Salgado, P. Salgado-Rebolledo, O. Valdivia, Phys. Lett. B 728 (2014) 99–104.}



 \bibitem{danhuaechf} \href{https://doi.org/10.1016/j.physletb.2025.139471}{D. H. Song, Phys. Lett. B 865, (2025) 139471.} 
 

 
 \bibitem{algebra}\href{ https://doi.org/10.48550/10.1007/BF01213402}{J. Ma\~nes, R. Stora, B. Zumino, Commun. Math. Phys. 102 (1985) 157.}
 \bibitem{seperat}\href{https://doi.org/10.1007/s11005-007-0148-0}{F. Izaurieta, E. Rodr\'iguez, P. Salgado, Lett. Math. Phys. 80 (2007) 127.}
 \bibitem{tansi}\href{https://doi.org/10.1088/1126-6708/2006/02/067}{P. Mora, R. Olea, R. Troncoso, J. Zanelli, JHFP. 02 (2006) 067.}
  \bibitem{mengy}\href{
     https://doi.org/10.48550/arXiv.2603.27588
 }{M. Y. Wu, D. H. Song, J. Yang, arXiv. 2603.27588 }
 
 
\end{thebibliography}
\end{document}